\documentclass{article}
\usepackage{amsmath}
\usepackage{amssymb}
\usepackage{theorem}
\usepackage{xspace}
\usepackage[hyphens]{url}
\usepackage{datetime}
\usepackage[usenames,dvipsnames,svgnames]{xcolor}
\usepackage{siunitx}
\usepackage{microtype}
\usepackage{authblk}
\usepackage[latin1]{inputenc}
\usepackage[margin=4cm]{geometry}

\usepackage{algorithmic, algorithm}

\usepackage{multirow}
\usepackage{slashbox}

\usepackage{epstopdf}
\usepackage{tikz}
\usetikzlibrary{shapes,arrows}
\usetikzlibrary{decorations,calc}

\newcommand{\bv}{\vspace{-.01in}}

\newcommand{\ic}{\ensuremath{\iota_C}}
\newcommand{\id}{\ensuremath{\iota_D}}
\newcommand{\is}{\ensuremath{\iota_S}}

\newcommand{\RR}{\ensuremath{\mathbb{R}}}
\newcommand{\RPP}{\ensuremath{\left]0,+\infty\right[}}
\newcommand{\RX}{\ensuremath{\left]-\infty,+\infty\right]}}
\newcommand{\KK}{\ensuremath{\mathbb{K}}}

\newcommand{\bR}{\ensuremath{\mathbf{R}}}
\newcommand{\bh}{\ensuremath{\mathbf{h}}}

\newcommand{\I}{\ensuremath{\operatorname{I}\xspace}}

\newcommand{\Diag}{\ensuremath{\operatorname{Diag}}}

\graphicspath{{./Figures/}}

\title{A constrained-based optimization approach for seismic data recovery problems}
\author[1,3]{Mai Quyen Pham\thanks{mai-quyen.pham@ifpen.fr}}
\author[2]{Caroline Chaux\thanks{caroline.chaux@latp.univ-mrs.fr}}
\author[1]{Laurent Duval\thanks{laurent.duval@ifpen.fr}}
\author[3]{Jean-Christophe Pesquet\thanks{pesquet@univ-mlv.fr}}
\affil[1]{IFP Energies nouvelles, Dir. Technologie\\ 
1 et 4 Av de Bois-Préau, 92852 Rueil-Malmaison Cedex - France}
\affil[2]{Aix-Marseille Univ., LATP UMR CNRS 7353\\ 
39 Rue F. Joliot-Curie, 13453 Marseille Cedex 13 - France\\ }
\affil[3]{Univ. Paris-Est, LIGM UMR CNRS 8049\\ 
5 Bd Descartes, 77454 Marne-la-Vall\'ee - France}

\renewcommand\Authands{ and }

\begin{document}

\maketitle

\begin{abstract}
Random and structured noise both affect seismic data, hiding the reflections of interest (primaries) that carry meaningful geophysical interpretation. When the structured noise is composed of multiple reflections, its adaptive cancellation is obtained through time-varying filtering, compensating inaccuracies
in given approximate templates. The under-determined problem can then be formulated as a convex optimization one, providing estimates of both filters and primaries.
Within this framework, the criterion to be minimized mainly consists of two parts: a data fidelity term and hard constraints modeling a priori information. This formulation   may avoid, or at least facilitate, some parameter determination tasks, usually difficult to perform in inverse problems. Not only classical constraints, such as sparsity, are considered here,  
but also constraints  expressed through hyperplanes, onto which the projection is easy to compute.
The latter constraints lead to improved performance by further constraining the space of geophysically sound solutions.
\end{abstract}


\section{Introduction}
\label{sec:intro}

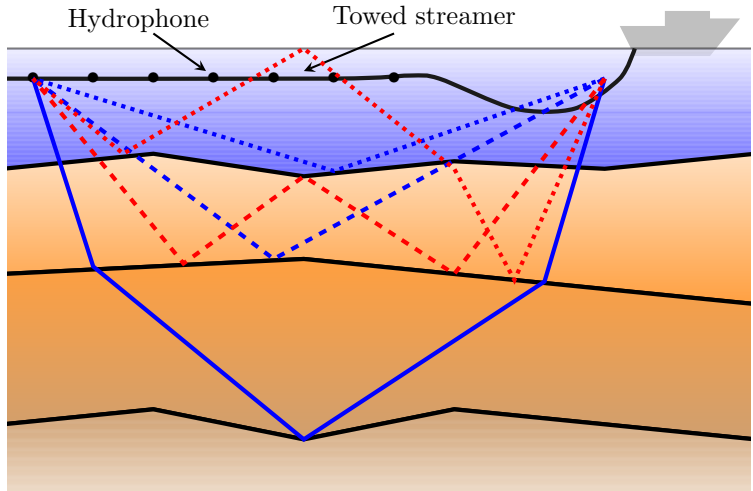
\begin{figure}[htb]
\begin{center}
\begin{tikzpicture}[scale=2]
\begin{scope} 
\clip ($(0,0)+2*(\pgflinewidth,\pgflinewidth)$) rectangle ($(5,3.5)-2*(\pgflinewidth,\pgflinewidth)$);
\fill   [xshift=0.2cm, fill=Silver]  (4,2.95) --  (4.5,2.95)  --  (4.7,3.20)  --  (4.50,3.20)  --  (4.50,3.25)  --  (4.20,3.25)  -- (4.20,3.15)  --  (3.9,3.15) -- cycle;
\draw[thick,bottom color=blue!90,top color = blue!10,opacity=.5] (0,2.2) -- (1,2.3) -- (2,2.15) -- (3,2.25) -- (4,2.2) -- (5,2.3) -- (5,3) -- (0,3) -- cycle;
\draw [ultra thick,black!90, xshift=0.2cm] plot [smooth, tension=0.5] coordinates { (4.0,3.0) (3.9,2.8) (3.6,2.6) (3.2,2.6) (2.7,2.8)  (2.5,2.82) (2,2.8) (-0.4,2.8)};
\draw (0.2,2.8) node {$\bullet$} (0.6,2.8) node {$\bullet$} (1.0,2.8) node {$\bullet$} (1.4,2.8) node {$\bullet$} (1.8,2.8) node {$\bullet$} (2.2,2.8) node {$\bullet$} (2.6,2.8) node {$\bullet$};
\draw (2.8,3.1) node[above] {Towed streamer} ; 
\draw [>=stealth,->,thick] (2.6,3.1) -- (2,2.85) ; 
\draw (0.9,3.05) node[above] {Hydrophone} ; 
\draw [>=stealth,->,thick] (1.0,3.1) -- (1.35,2.85) ; 
\draw[ultra thick,bottom color=orange!75,top color = orange!25]  (0,2.2) -- (1,2.3) -- (2,2.15) -- (3,2.25) -- (4,2.2) -- (5,2.3) --  (5,1.3) -- (2,1.6) -- (0, 1.5) --  cycle;
\draw[ultra thick,bottom color=brown!75,top color = orange!75] (0, 1.5) --  (2,1.6) --  (5,1.3) --  (5,0.4) -- (4,0.5) -- (3,0.6) -- (2,0.4) --   (1,0.6) -- (0,0.5) -- cycle;
\draw[ultra thick,bottom color=brown!25,top color = brown!75] (0,0.5) -- (1,0.6) -- (2,0.4) -- (3,0.6) -- (4,0.5) -- (5,0.4) -- (5,0) -- (0,0) -- cycle;
\draw [ultra thick,  color=blue, dotted] (4,2.8) -- (2.2,2.185)-- (0.2,2.8);
\draw [ultra thick,  color=blue, dashed] (4,2.8) -- (1.8,1.6)-- (0.2,2.8);
\draw [ultra thick,  color=blue] (4,2.8) -- (3.6,1.45)-- (2.0,0.4)--  (0.6,1.55)-- (0.2,2.8);
\draw [ultra thick,  color= red!100, dashed] (4,2.8) -- (3.0,1.5)-- (2.0,2.15) -- (1.2,1.57) -- (0.2,2.8);
\draw [ultra thick,  color= red!100, dotted] (4,2.8) -- (3.4,1.46)-- (3,2.2) -- (2,3) -- (0.8,2.3) -- (0.2,2.8);

\end{scope}
\end{tikzpicture}
\end{center}
\vspace{-0.3cm}
\caption{Principles of marine seismic data acquisition and wave propagation. Towed streamer with hydrophones. Reflections on
different layers (primaries in  blue), and  reverberated disturbances (multiple in dotted and dashed red).\label{fig_marine_seismic_reflection_multiple}}
\vspace{-0.4cm}
\end{figure}
Adaptive filtering techniques are meant to  optimize coefficients of variable filters, according to adapted cost functions working on error signals. Adaptive subtraction \cite{Liu_S_2009_misc_progress_rsrmeiasf,Ventosa_S_2012_j-geophysics_adaptive_mswbcuwf} is
  at play in seismic data recovery problems where approximate models  are adapted or matched to actual data, throughout adaptive filters. These models are obtained from geophysical modeling, and known a priori. One such situation is the filtering of secondary reflexions, or multiples.  Geophysical signals of interest, named primaries, follow wave paths depicted in dotted, dashed and solid blue in Fig. \ref{fig_marine_seismic_reflection_multiple}. Since the data recovery problem is generally under-determined, geophysicists have developed pioneering sparsity-promoting techniques. For instance, robust, $\ell_1$-promoted deconvolution \cite{Claerbout_J_1973_j-geophysics_robust_med} or complex wavelet transforms \cite{Morlet_J_1982_j-geophysics_wave_pstp1cssmm,Morlet_J_1982_j-geophysics_wave_pstp2stcw} still pervade many areas of signal processing. 
Although the contributions are generally considered linear, several types of disturbances, structured or more stochastic, affect the relevant information present in seismic data. 
Multiples correspond to seismic waves bouncing between layers \cite{Essenreiter_R_1998_j-ieee-tsp_multiple_rasdb}, as illustrated with red dotted and dashed lines in Fig. \ref{fig_marine_seismic_reflection_multiple}. These reverberations share waveform and frequency contents similar to primaries, with longer propagation times. 
From the standpoint of geological information interpretation, they often bedim deeper target reflectors. For instance, the dashed-red multiple path may possess a total travel time comparable with that of the solid-blue primary. Their separation is thus required for accurate subsurface characterization. We suppose here that one or several approximate templates of  potential  multiples are determined, off-line, based on primary reflections identified in above layers or wave-propagation modeling. Model-based multiple filtering is similar  to   adaptive echo cancellation practice (see \cite{Pham_M_2014_PREPRINT_primal-dual_proximal_astbafasmr} for details), and is now considered  as a geophysics industry standard.

We propose a methodology for primary/multiple adaptive separation based on these approximate templates. It addresses at the same time  structured reverberations and a more stochastic part. Let $n\in \{0,\ldots,N-1\}$ denote the time index for the observed seismic trace $z$, acquired by a given sensor (here, an hydrophone). We assume, as customary in seismic, a linear model of contributions:
\begin{equation}
z^{(n)} = \overline{y}^{(n)} + \overline{s}^{(n)} +  b^{(n)}\,.
\label{eq:model}
\end{equation} 
The unknown signal of interest (primary, in blue) and the sum of undesired, secondary reflected signals (different multiples, in red) are denoted, respectively, by $\overline{y}=(\overline{y}^{(n)})_{0 \le n < N}$ and $(\overline{s}^{(n)})_{0 \le n < N}$. Other unstructured contributions are gathered in the noise term $b= (b^{(n)})_{0 \le n < N}$. 
We assume that $J$ templates $(r^{(n)}_j)_{0 \le n < N, 0 \le j < J}$ for the disturbance signal are available, which are related to $(\overline{s}^{(n)})_{0\le n< N}$ through an FIR (Finite Impulse Response), possibly non-causal, linear model
\begin{equation}\label{e:sn}
\overline{s}^{(n)} = \sum_{j=0}^{J-1}\, \sum_{p=p'}^{p'+P_j-1} \overline{h}_j^{(n)}(p) r_j^{(n-p)}
\end{equation}
where $\overline{h}_j^{(n)}$ is an unknown impulse response ($P_j$ tap coefficients) corresponding to template $j$ and time $n$ and where $p' \in \{-P_j+1,\ldots,0\}$ ($p'=0$ corresponds to the causal case). It must be emphasized that the dependence w.r.t. the time index $n$ of the impulse responses implies that the filtering process is not time invariant, although it can be assumed slowly varying in practice.

The purpose of this work is to provide means to identify $\overline{y}$ and $\overline{h}_j$, by imposing hopefully meaningful constraints onto the above system.

\section{Relation to prior work}
\label{sec:related}

The separation of primaries and multiples is a classical issue in seismic exploration. Most published solutions, tailored to specific levels of prior knowledge, are very dependent on seismic data-sets. They generally rely on adapted transforms (Radon, Fourier transforms) and some form of least-squares adaptive filtering. 
Among the vast literature, we  refer to \cite{Ventosa_S_2012_j-geophysics_adaptive_mswbcuwf,Song_J_2013_j-appl-geophys_comparing_tfimem}, for a recent account on adaptive  subtraction of multiples, including shortcomings of standard $\ell_2$-based methods. 
With weak primary/multiple decorrelation, poor data stationarity or higher noise levels, traditional methods fail. Due to the parsimonious layering \cite{Walden_A_1986_j-geophys-prospect_nature_ngprcsd} of the subsurface (illustrated in Fig. \ref{fig_marine_seismic_reflection_multiple}), sparsity promotion suggests the use of sparsifying transforms (e.g. wavelet/curvelet frames \cite{Herrmann_F_2004_p-cseg_curvelet_ipame,Neelamani_R_2010_j-geophysics_adaptive_sucvct}), potentially combined with robust norms (approximate $\ell_1$ in \cite{Guitton_A_2004_j-geophys-prospect_ada_smul1n}), quasi-norms or source separation methods \cite{Donno_D_2011_j-geophysics_improving_mrulsdfica,Duarte_L_2012_p-eusipco_seismic_wsmrpca}. To date, their genericity may be limited by the number of possible penalties to constrain feasible solutions, and the crucial issue of hyperparameter determination in such methods.

In \cite{Pham_M_2014_PREPRINT_primal-dual_proximal_astbafasmr,Pham_M_2013_p-icassp_seismic_mrpdpa}, 
 the authors 
incorporate
plausible knowledge via additional metrics. Prior multiple templates are supplemented with Gaussian noise assumptions, wavelet-domain sparsity, smooth  variations and energy concentration criteria.
Joint estimation of primaries and adaptive filters is performed with a proximal algorithm. To alleviate the hyperparameter estimation issue, we reformulate the previous approach as a constrained minimization problem. This allows us to more easily determine data-based parameters. We focus here on the exploration of various constraint efficiency in wavelet frame subbands for the primary signal.
Interestingly, 
convex sets defined as appropriate hyperplanes can outperform standard $\ell_1$-ball constraints.

The paper is organized as follows. In Section \ref{sec:var-formulation} we rewrite the observation model and formulate the constrained optimization problem. The definition of the constraint sets and the adopted optimization strategy follows. 
Section \ref{sec:results} details the simulation results. Finally, conclusions are drawn in Section~\ref{sec:conclusions}.

\section{Constrained formulation}
\label{sec:var-formulation}

\subsection{Observation model}
Model \eqref{e:sn} can be written more concisely as
\begin{equation}
\overline{s}=\bR\overline{\bh}
\end{equation} 
by appropriately defining $\bR \in \RR^{N \times Q}$, where $Q=NP$ with $P=\sum_{j=0}^{J-1} P_j$ and $\overline{\bh} \in \RR^Q$ \cite{Pham_M_2013_p-icassp_seismic_mrpdpa}.
On the one hand, the matrix $\bR$ contains the $J$ templates for every time index $n$ and tap index $p$. On the other hand, the vector $\overline{\bh}$ is similarly defined as the concatenation of all (unknown) time-varying filter impulse responses.
With this notation, the observed data $z$ are given by
\begin{equation}
z = \overline{y} + \bR\overline{\bh} + b.
\label{eq:model_gen}
\end{equation}
Now, we  turn our attention to solving the ill-posed inverse problem of estimating $\overline{y}$ and $\overline{\bh}$ from the observation vector $z$.

\subsection{Constrained problem formulation}
Our objective here is to propose a variational approach aiming at providing relevant estimates of the primary signal $\overline{y}$ and time-varying filters $\overline{\bh}$ related to multiples.
To this end, define an objective function composed of two convex terms being related to either $\overline{y}$ or $\overline{\bh}$ through functions $\varphi : \RR^N \to \RX$ and $\rho \colon \RR^Q \to \RX$, respectively. 
We propose to solve the following constrained minimization problem
\begin{align}
& \underset{y\in \RR^N,\, \bh \in \RR^Q}{\text{minimize}}\quad \alpha \rho(\bh) + (1-\alpha) \varphi(y) \nonumber \\
& \text{subject to} \; \begin{cases} 
\psi(z-y-\bR\bh) \leq 1 \\
\bh \in C \\  
Fy \in D \end{cases}
\label{eq:fct_constrain}
\end{align}
where $\alpha \in [0,1]$, $F \in \RR^{K\times N}$, $K \geq N$,  models a (non necessarily tight) frame operator  \cite{Jacques_L_2011_j-sp_panorama_mgrisdfs}, and  $C$ and $D$ are nonempty closed convex constraint sets that are defined hereafter.

\subsection{Constraint set definitions}
We discuss here the different choices that can be adopted for the potential functions as well as the convex sets $C$ and $D$. These choices reflect some a priori knowledge one may have on the variables to be estimated. The idea in addressing a constrained formulation instead of a regularized formulation is to avoid or, at least to facilitate, hyperparameter estimation. This is detailed later on, in the simulation part.

\subsubsection{Coupling constraint: function $\psi$}
The seismic noise $b$ 
is naturally assumed to be additive white Gaussian with zero-mean and variance $\sigma^2$. A natural choice for $\psi$ is thus to take $\psi=\| \, . \,\|^2/(N\sigma^2)$. 
When the noise variance is unknown, it can be easily and accurately estimated by using classical techniques such as the median absolute deviation (MAD) \cite{Hampel_F_1974_j-asa_influence_crre} wavelet estimator \cite[p. 446]{Donoho_D_1994_j-biometrika_ideal_saws}.

\subsubsection{Hard constraints on  filters $\bh$: convex set $C$}
As mentioned earlier, the filters are assumed to be time-varying. However, real case study showed that those filters have smooth variations along time index $n$. To ensure that this a priori characteristic is satisfied for the estimated filters, we propose to introduce the following upper bound on the impulse response variations \cite{Pham_M_2013_p-icassp_seismic_mrpdpa}:
\begin{equation}
\forall (j,p,n), \qquad |h_j^{(n+1)}(p) - h_j^{(n)}(p)| \le \varepsilon_{j,p}
\label{eq:convex_C}
\end{equation}
where $\varepsilon_{j,p} \in \RPP$.

\subsubsection{Hard constraints on  primaries  $y$: convex set $D$}
\label{ssec:convexD}
First of all, the primary signal $y$ is assumed to be well represented onto a wavelet frame \cite{Jacques_L_2011_j-sp_panorama_mgrisdfs}, whose analysis operator is $F \in \RR^{K\times N}$. 
To further account for the wavelet analysis frame coefficient properties, we propose to split the convex set $D$ as $D_1 \times \cdots \times D_{\mathcal{L}}$. Indeed, the idea here is to construct $\mathcal{L}$ partitions of $\{1, \ldots ,K\}$ denoted by $\{\KK_\ell \mid \ell \in \{1,\ldots,\mathcal{L}\}\}$ where $\mathcal{L}$ corresponds to the number of subbands and $\KK_\ell$ is the $\ell$-th subband.
In this work, we investigate two kinds of convex sets $\left(D_{\ell}\right)_{\ell \in \{1,\ldots,\mathcal{L}\}}$:
\begin{enumerate}
\item The first one is widely used in the literature and consists of defining sets of the form: for every $\ell \in \{1,\ldots,\mathcal{L}\}$,
$
D_{\ell} = \{(x_k)_{k\in \KK_\ell} \mid \sum_{k\in \KK_\ell} \phi_\ell(x_k) \leq \eta_\ell \}, 
$
where $\eta_\ell \in \RR$, and $\phi_\ell: \RR^{|\KK_\ell|} \to \RX$ is a proper lower-semicontinuous convex function. 
For example, this constraint set definition enables to incorporate sparsity constraints on the wavelet frame coefficients
in the optimization problem,  by choosing e.g. $\phi_\ell = |\cdot|$.
\item The second one is more original and consists of defining hyperplanes: for every $\ell \in \{1,\ldots,\mathcal{L}\}$, 
$D_{\ell} = \{(x_k)_{k\in \KK_\ell} \mid \sum_{k\in \KK_\ell} \phi_\ell((FLz)_k) x_k = \eta_\ell \},$ where $L \in \RR^{N\times N}$ is an appropriate linear operator and $\phi_\ell : \RR \to \RR$.
The simplest choice for $L$ is to take the identity operator $L=\I$. An alternative choice, which is reminiscent of Wiener filtering, is
\begin{multline}
L=\lambda_1 \Diag\Big( (1 + \lambda_1+\lambda_2\|\bR^{(0)}\|^2)^{-1} \; , \ldots ,(1 + \lambda_1 +\lambda_2\|\bR^{(N-1)}\|^2)^{-1} \Big)
\label{eq:opL}
\end{multline}
where $(\lambda_1,\lambda_2) \in \RPP^2$ and for every $n \in\{0,\ldots,N-1\}$, $\bR^{(n)}$ denotes the $n$-th row of matrix $\bR$.
\end{enumerate}

\subsection{Optimization strategy}
\label{ssec:algo}

One can note that Problem \eqref{eq:fct_constrain} can be reexpressed as
\begin{equation}
\underset{y\in \RR^N,\, \bh \in \RR^Q}{\text{minimize}} f\left( y , \bh \right)  + \is\left(z-[\I\;\bR] \begin{bmatrix} y \\ \bh \end{bmatrix}\right) + \ic(\bh) + \id(Fy)
\label{eq:fct_regul}
\end{equation}
where 
$f \colon \RR^N\times \RR^Q \to \RX \colon (u,v) \mapsto \alpha \rho(v)+ (1-\alpha) \varphi(u)$,\\ 
$S=\left\{w \in \RR^N \mid \|w\|^2 \leq N\sigma^2\right\}$ and $\is$ is the indicator function of the set $S$ defined as
\begin{equation}
\is(u)= \begin{cases} 0  & \text{if } u\in S\\
 +\infty & \text{otherwise}
\end{cases}
\end{equation}
(a similar notation being used for $C$ and $D$).
Such convex optimization problems, involving the sum of $4$ convex functions and various linear operators, can be solved in an efficient manner by using primal-dual approaches such as the Monotone+Lipschitz Forward-Backward-Forward (M+L FBF) algorithm \cite{Combettes_P_2012_j-set-valued-var-anal_primal_dsasimclpstmo} as well as the algorithm  in
\cite{Chambolle_A_2011_j-math-imaging-vis_first_opdacpai}, which was recently extended in \cite{Vu_B_2013_j-adv-comput-math_splitting_admiico,Condat_L_2013_j-optim-theory-appl_primal-dual_smcoilplct}.
The functional to be minimized being composed of convex functions as well as indicator functions of convex sets, the algorithm  typically requires to compute, in parallel, proximal operators and projections onto the different closed convex sets.
Concerning proximity operators, closed-form expressions for a wide class of convex functions can be found in \cite{Combettes_P_2011_incoll_proximal_smsp}. The projection onto $C$ is explicit and reduces to projections onto hyperslabs (after appropriate splitting). Similarly, when considering  affine constraint for convex set $D$ (second case) the projection is explicit. For all the remaining cases, projections onto $\ell_p$-ball are performed, some of which can be computed explicitly (e.g. $\ell_2$-ball or $\ell_\infty$-ball) or iteratively (e.g. $\ell_1$-ball \cite{Friedlander_M_2011_p-siam-optim_spa_oaa}).

\section{Results}
\label{sec:results}

\begin{figure}[htb]
\centering
\includegraphics[height=7cm]{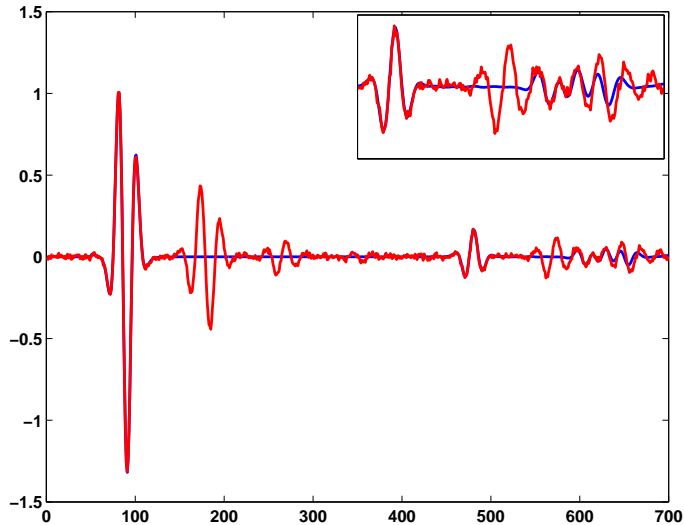}
\caption{Observed signal $z$ (red; $\sigma=0.01$), original  $y$ (blue). \label{fig:estims_001a}}
\vspace{-0.3cm}
\end{figure}

\begin{figure}[htb]
\centering
\includegraphics[height=7cm]{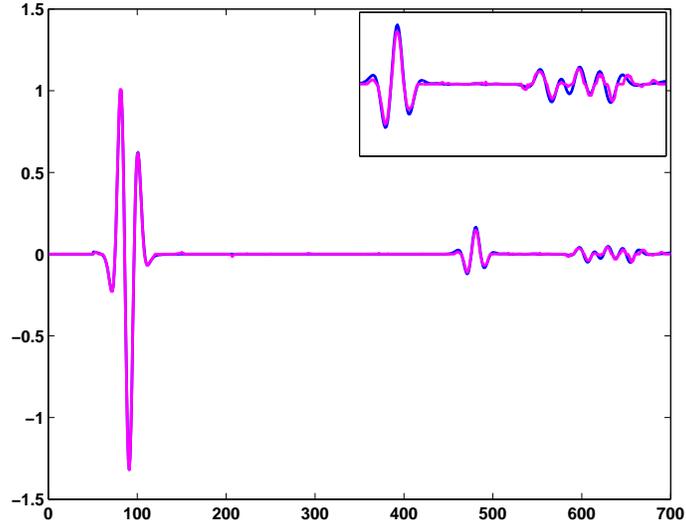}
\caption{Estimated signal $\hat{y}$ (magenta), original signal $y$ (blue). $D$ is the intersection of two hyperplanes defined from the identity and the sign functions. \label{fig:estims_001b}}
\vspace{-0.3cm}
\end{figure}

\begin{figure}[htb]
\centering
\includegraphics[height=7cm]{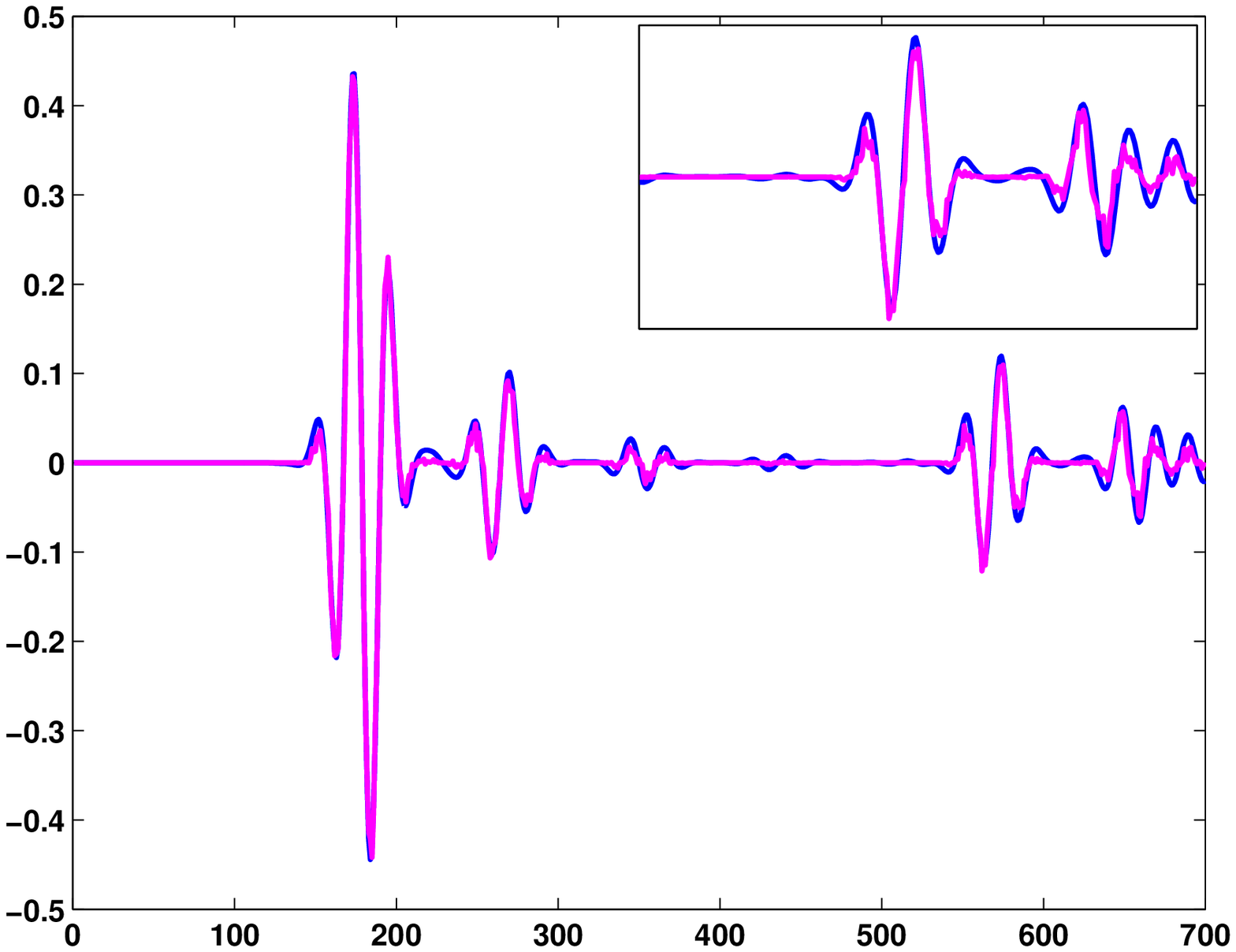}
\caption{Estimated multiples $\hat{s}$ (magenta), original multiples $s$ (blue). $D$ is the intersection of two hyperplanes defined from the identity and the sign functions. \label{fig:estims-s_001}}
\vspace{-0.3cm}
\end{figure}

Simulation tests are performed on synthetic seismic data. From realistic primary signal $\overline{y}$ and templates $\bR$ where $J=2$, we generated observations according to model \eqref{eq:model_gen} where appropriate time-varying filters $\overline{\bh}$ are used with $P_1=6$ and $P_2=6$. 
The primary signal as well as the observations with $\sigma=0.01$ are represented  Fig. \ref{fig:estims_001a}.
The criterion to be minimized is defined by \eqref{eq:fct_constrain} 
where $\varphi$ is chosen to be the $\ell_1$-norm, and $\rho$ is the squared $\ell_2$-norm (see \cite{Pham_M_2014_PREPRINT_primal-dual_proximal_astbafasmr} for more extensive tests on different choices for $\rho$).
Concerning the constraint set definitions, on the one hand, $C$ is defined by \eqref{eq:convex_C} where, for every $p$, $\varepsilon_{1,p}= \varepsilon_{2,p}= 0.17$. On the other hand, $D$ is defined by choosing $F$ to be a (non tight) undecimated wavelet frame with 
Daubechies wavelets of length $8$ and $4$ resolution levels. 
We have considered the two possibilities described in Section \ref{ssec:convexD} where, in the first case (inequality constraint), $\phi_\ell \equiv \phi$ where $\phi$ is either the $\ell_1$, $\ell_2$  or the $\ell_\infty$-norm. In the second case, $\phi_\ell \equiv \phi$ where $\phi$ is either the identity  or the sign function; furthermore, $L$ is chosen according to \eqref{eq:opL} where $\lambda_1 = 0.02$ and $\lambda_2 = 0.001$. In this last case, both affine constraints have also been considered jointly (intersection of the two constraint sets).

\begin{table}[htb]\centering
\resizebox{9cm}{!}{\begin{tabular}{|c|c|c|c|c|c|c|}
\hline
  $\sigma$ & \multicolumn{3}{c|}{$0.01$} & \multicolumn{3}{c|}{$0.04$}\\
\hline
 $\phi$ & $  \alpha  $ &  SNR$_y  $  &  SNR$_s$ & $\alpha$ & SNR$_y$ & SNR$_s$ \\
\hline
 0 &0.4& 23.98 &15.79&0.9& 15.03&  9.60\\
\hline
$\ell_1$&  0.4  & 25.98& 16.16 & 0.9  &   18.19 & 6.61\\ 
\hline
$\ell_2$ &  0.6  &  25.59  & 16.02 & 0.8  &   17.84 &   9.20\\
\hline
$\ell_{\infty}$ &0.6  &  24.48  & 15.81& 0.8  &   16.24 & 8.69\\
\hline
\hline
$\I$ &0.4  &   26.19  & 15.81& 0.2& \textbf{19.74} & 8.84\\ 
\hline
$\text{sign}$ &0.3  & 24.43 &  14.73 & 0.1 &  14.75 & 4.58\\
\hline
$\I + \text{sign}$ &0.3  &  \textbf{26.40} &  15.56 & 0.1&   18.43 & 5.94\\
\hline 
\end{tabular}}
\caption{SNR for the estimations of $y$ and $s$ (SNR$_y$ and  SNR$_s$, resp.) in \si{dB} considering
different convex constraint sets $D$ and two noise levels. Upper table part: ``classical constraints'' and lower table part: hyperplane contraints. \label{tab:snr}}
\end{table}

Restoration results, using M+L FBF algorithm, for the primary signal in the case when $\sigma=0.01$, are displayed 
in  Fig. \ref{fig:estims_001b}. 
The associated estimated multiples are plotted in Fig.~\ref{fig:estims-s_001}. From these two figures, one can note that the multiples are quite
well estimated and adequately separated from the primary. 
The stochastic part 
is accurately removed, even if
 some residual noise remains, for instance  when the signal is of
small amplitude.
Table \ref{tab:snr} shows the signal-to-noise ratios obtained for the estimation of $y$ and $s$. Simulations have been run for different convex sets $D$ and for two noise levels (with standard-deviation $\sigma = 0.01$ and  $\sigma = 0.04$).  The notation $\phi=0$ has been used in the case when no constraint
is applied to $Fy$. This allows us to evaluate the gain  (up to $1.4$ dB)
brought by 
the introduction of prior information on $Fy$ 
through a constrained 
formulation.

\section{Conclusions}
\label{sec:conclusions}

This paper focuses on the constrained convex formulation of adaptive multiple removal.
The proposed approach, based on proximal methods, is quite flexible and allows us to integrate a large panel of hard constraints corresponding to a priori knowledge on the data to be estimated (i.e. primary signal and time-varying filters). A key observation is that some of the related constraint sets can be expressed through hyperplanes, which are not only more convenient to design, but also easier to implement through straightforward projections. Since sparsifying transforms and  constraints strongly interact \cite{Pham_M_2014_PREPRINT_primal-dual_proximal_astbafasmr}, we now  study the class of hyperplane constraints of interest as well as their inner parameters, together with the extension to higher dimensions.

\end{document}